$^{59}$Co NQR Study on Superconducting Na$_x$CoO$_2 \cdot y$H$_2$O

Yoshiaki Kobayashi, Mai Yokoi and Masatoshi Sato

*Department of Physics, Division of Material Science, Nagoya University, Furo-cho, Chikusa-ku, Nagoya 464-8602, Japan*

Layered Co oxide Na$_x$CoO$_2 \cdot y$H$_2$O with a superconducting transition temperature $T_c$ =4.5 K has been studied by $^{59}$Co NQR. The nuclear spin relaxation rate $1/^{59}T_1$ is nearly proportional to temperature $T$ in the normal state. In the superconducting state, it exhibits the coherence peak and decreases with decreasing $T$ below ~0.8$T_c$. Detailed comparison of the $1/T_1T$ values and the magnetic susceptibilities between Na$_x$CoO$_2 \cdot y$H$_2$O and Na$_x$CoO$_2$ implies that the metallic state of the former system is closer to a ferromagnetic phase than that of the latter. These experimental results impose a restriction on the mechanism of the superconductivity.

corresponding author: M. Sato (e-mail: e43247@nucc.cc.nagoya-u.ac.jp)

KEYWORDS: Na$_x$CoO$_2 \cdot y$H$_2$O, superconductivity, NQR

Co oxides, which consist of linkages of $CoO_6$ octahedra, have attracted much attention because they exhibit various notable physical properties. $RCoO_3$ (R=Y and various rare earth elements) and other related systems with the perovskite structure are examples of such systems.[1-5] In such systems, the electron occupation of the lower $t_{2g}$ and higher $e_g$ bands of Co ions changes in a rather significant way when the temperature $T$ is changed or other structural parameters are changed. Due to the existence of the strong Hund coupling between these electrons in different bands, the double exchange type coupling becomes effective with increasing number of electrons excited to the $e_g$ band. Then, the systems are considered to be in the proximate region of a ferromagnetic phase.

$Na_xCoO_2$ is another example of such systems. A two-dimensional triangular lattice of Co is formed in the oxide by the edge-sharing $CoO_6$ octahedra. $Na_xCoO_2$ has a high thermoelectric performance,[6-8] and is also considered to be in the vicinity of a ferromagnetic phase.[9,10] Recently, Takada et al.[11] have found superconductivity with a transition temperature $T_c$ of ~5 K for $Na_xCoO_2 \cdot yH_2O$ prepared by deintercalating $Na^+$ and intercalating $H_2O$ into $Na_xCoO_2$, and various kinds of experimental and theoretical studies have been carried out[11-19] to investigate the mechanism of the superconductivity.

In the present work, the $^{59}Co$ nuclear spin-lattice relaxation rate ($1/T_1$) has been measured for a sample of $Na_xCoO_2 \cdot yH_2O$ with $T_c$ = 4.5 K in a zero magnetic field by $^{59}Co$ NQR up to 30 K, and based on the observed results, the pairing symmetry is discussed. The normal state data are compared with those of the parent compound $Na_{0.75}CoO_2$.

A powder sample of $Na_xCoO_2 \cdot yH_2O$ was synthesized from $Na_{0.75}CoO_2$ as described in ref. 11 and confirmed to be the single phase by X-ray diffraction measurements. The lattice parameters $a$ and $c$ are 2.8245 and 19.7101 Å, respectively. The uniform magnetic susceptibility $\chi$ was measured using a SQUID magnetometer with magnetic fields $H$= 5 Oe and 1 T. The data for $H$= 5 Oe are shown in Fig. 1 as a function of $T$, where the onset of the superconductivity is found at $T$ ~4.5 K. The magnitudes of diamagnetic signals taken at $T$ =2.0 K under the conditions of zero-field cooling and field cooling correspond to the volume fractions of ~24% and ~6%, respectively. These values are rather large for the powder sample with two-dimensional nature and therefore exclude the possibility that the superconductivity is filamentary. The $T_1$ of Co nuclei was measured by detecting nuclear magnetization after applying an inversion pulse in the frequency region around $3\nu_Q$ (the 5/2→7/2 transition), $\nu_Q$ being the electrical quadrupolar frequency (~4.2 MHz at 4.23 K). This value of $\nu_Q$ is consistent with the one estimated from the $^{59}Co$ NMR spectra of this sample.[20] The nuclear magnetization $m$ measured as a function of time $t$ elapsed after the inversion pulse, was analyzed under the assumption that there exist two components with shorter and longer relaxation times, $T_{1short}$ and $T_{1long}$, respectively. The following theoretical equation[21]

$$[m(\infty)-m(t)]/m(\infty) = 2\{c[0.204\exp(-3t/T_{1short})+0.649\exp(-10t/T_{1short})+0.136\exp(-21t/T_{1short})] + (1-c)[0.204\exp(-3t/T_{1long})+0.649\exp(-10t/T_{1long})+0.136\exp(-21t/T_{1long})]\} \quad (1)$$

with the number fraction $c$ of nuclear spins which have a shorter relaxation time $T_{1\text{short}}$, was used. A typical example of the fittings to the observed $m$-$t$ curves is shown in the inset of Fig. 2. The $c$ values are found to be ~(0.96-0.98). $T_{1\text{long}}$ is of the order of 1 s and the origin of this longer relaxation time may be due to the existence of a small amount of impurity phase(s). Because $T_{1\text{short}}$ is smaller than $T_{1\text{long}}/10$ even in the lowest temperature region, both kinds of $T_1$ can be reliably estimated. Hereafter, $T_1$ denotes the relaxation time $T_{1\text{short}}$ corresponding to the main phase.

The main panel of Fig. 2 shows the $T$-dependence of the nuclear spin relaxation rate divided by $T$, ($1/T_1T$) up to 30K. In the normal state, the relation $1/T_1T \cong 13$ s$^{-1}$K$^{-1}$ holds approximately down to $T_c$. This nearly $T$-independent behavior of $1/T_1T$ is expected for conventional metallic systems and should be contrasted with the rather significant $T$-dependent behavior of the uniform susceptibility $\chi$ observed for the same sample as was used in the NQR measurements (see the data of $\chi$ shown in Fig. 3). This difference of the $T$-dependences between $1/T_1T$ and $\chi$ of the sample Na$_x$CoO$_2 \cdot y$H$_2$O indicates that the $T$-dependent part of $\chi$ is not intrinsic to Na$_x$CoO$_2 \cdot y$H$_2$O. In the superconducting state, $1/T_1T$ increases just below $T_c$ and then decreases with decreasing $T$. The maximum is at ~$0.8T_c$. It is obvious that the small but clear coherence peak exists just below $T_c$. It indicates that a finite energy gap exists throughout the Fermi surface, which stands in clear contrast to the point node gap observed in high-$T_c$ copper oxides.

In Fig. 4, the logarithm of $T_1$ is plotted against $1/T$. The solid curve in the figure exhibits the Korringa relation $1/T_1T$=constant. In the low-$T$ region, the $T$-dependence of the observed $1/T_1$ values is much weaker than the exponential-type $1/T$-dependence expected for the case of a finite gap. It is weaker than the $T^3$-dependence, too.

How can we understand the presently observed behaviors of $1/T_1$, the existence of the coherence peak and the weak $T$-dependence of $1/T_1$? To get a clue for answering these questions, we compare the results with those of superconductors with magnetic impurities.[22,23] La$_3$Al is considered to have a superconducting gap with $s$-wave symmetry.[22] For La$_{3-x}$Gd$_x$Al obtained by doping of magnetic Gd atoms, $T_c$ decreases to $0.6T_c(x=0)$ for $x$=0.02. For the sample, the $1/T_1$-$T$ curve of $^{27}$Al exhibits the coherence peak just below $T_c$. $1/T_1$ gradually decreases as $T$ decreases, where it has been considered that due to effects of the Gd magnetic moments, the nuclear spin relaxation rate does not exhibit an exponential-type decay at low temperatures. The similarity of the behaviors of $1/T_1$ of the present system to those observed for La$_{3-x}$Gd$_x$Al seems to indicate that the present system has an $s$-wave type energy gap and that exponential-type dependence of $1/T_1$ on $1/T$ cannot be observed because of the effects of a very small amount of magnetic moments which may exist in the sample. (It is noted here that judging from the $m$-$t$ curves, it is not likely that there exists severe inhomogeneity of the superconducting order parameter in the sample.)

For the nonintercalated sample Na$_{0.75}$CoO$_2$, we have also measured the $1/^{59}T_1$ at the peak position corresponding to the central transition of $^{59}$Co NMR.[9)] The values of $1/^{59}T_1$ were determined by fitting theoretical curves calculated by the following equation to the observed ones.

$[m(\infty)-m(t)]/m(\infty)$
$=2\{c[0.012\exp(-t/T_{1\text{short}})+0.068\exp(-6t/T_{1\text{short}})+0.206\exp(-15t/T_{1\text{short}})+0.714\exp(-28t/T_{1\text{short}})]$
$\quad +(1-c)[0.012\exp(-t/T_{1\text{long}})+0.068\exp(-6t/T_{1\text{long}})+0.206\exp(-15t/T_{1\text{long}})+0.714\exp(-28t/T_{1\text{long}})]\}$  (2)

The number fraction $c$ of nuclear spins with the short $T_{1\text{short}}$ was found to be 0.96-0.98 as in the case of Na$_x$CoO$_2 \cdot y$H$_2$O. In Fig. 5, the relaxation rate of Na$_{0.75}$CoO$_2$ is shown in the form of $1/T_1T$ -$T$, together with that of Na$_x$CoO$_2 \cdot y$H$_2$O, where the relation $1/^{59}T_1T \cong a+b/T$ ($a$ ~3/K/s and $b$ ~21/s) holds for Na$_{0.75}$CoO$_2$. As for the Knight shift, the isotropic part $K_{\text{iso}}$ of Na$_x$CoO$_2$ ($x$=0.5 ~0.75) exhibits very weak $T$-dependence and does not follow the $T$ dependence of $\chi$, while the anisotropic part follows the $T$ dependence of $\chi$.[9)] As was discussed in detail in ref. 9, these results indicate that the electron system of Na$_{0.75}$CoO$_2$ is essentially itinerant and the $b/T$ term can be considered to be the contribution from the Co sites around the localized Co moments, which are probably induced by lattice imperfections within the CoO$_2$ layers. The susceptibility corresponding to these itinerant electrons is deduced by subtracting the Curie term as shown by the chain line in Fig. 3. For Na$_x$CoO$_2 \cdot y$H$_2$O, the susceptibility of the itinerant electrons is similarly deduced and shown by the broken line in Fig. 3. The ratios of the susceptibilities of these itinerant electron systems of Na$_{0.75}$CoO$_2$ and Na$_x$CoO$_2 \cdot y$H$_2$O and the $T$-independent parts of $1/^{59}T_1T$ are both about 1 : 4. Since $1/T_1T$ depends on the square of the density of states $N(\varepsilon_F)$ at the Fermi level, the equal ratios obtained for the susceptibilities and for $1/^{59}T_1T$ cannot be explained if the effects of electron correlation are not considered. The results should be considered to indicate that the metallic state of Na$_x$CoO$_2 \cdot y$H$_2$O is closer to a ferromagnetic phase than that of Na$_{0.75}$CoO$_2$. (As was reported in ref. 10, Na$_x$CoO$_2$ itself is located in the proximity of a ferromagnetic phase.) The present results indicate that the superconductivity of Na$_x$CoO$_2 \cdot y$H$_2$O appears near a ferromagnetic phase and has $s$-wave symmetry.

We find, in $1/T_1T$ of Na$_x$CoO$_2 \cdot y$H$_2$O, a component proportional to $1/T$ (see Figs. 2 and 5). It may be a contribution from localized moments in the sample as in the case of Na$_{0.75}$CoO$_2$. Although the magnitude of the component is smaller than that observed for Na$_{0.75}$CoO$_2$, the rate of decrease of $1/T_1T$ with decreasing $T$ becomes larger in the temperature region well below $T_c$, when the component proportional to $1/T$ is subtracted from the observed values.

The experimental findings stated above present a rather severe restriction on the mechanism of the superconductivity. The existence of the coherence peak or the possible $s$-wave symmetry is, naively speaking, considered to be against the mechanisms predicting the $d$-wave or $p$-wave symmetry of the order parameter,[14-18)] though fully gapped order parameters are predicted even for the $d$-wave symmetry ($x^2$-$y^2$+$ixy$).[14-17)] The superconductivity predicted by Sano and Ono,[19)] who have studied strongly correlated electrons in two energetically separated bands and considered the Hund coupling between the electrons in these different bands, may be a candidate for the relevant mechanism. The present result that $Na_xCoO_2 \cdot yH_2O$ is located near a ferromagnetic phase also suggests the relevance of this mechanism, because their calculations predict the occurrence of superconductivity in the proximate region to a ferromagnetic phase. Of course, there exists the possibility that the superconductivity is just brought about by the electron-phonon coupling as in ordinary superconductors.

In conclusion, we have measured the nuclear spin relaxation rate $1/T_1$ of $Na_xCoO_2 \cdot yH_2O$. The existence of the coherence peak just below $T_c$ indicates that the system is a fully gapped superconductor. The very gradual decrease in $1/T_1$ in the low-temperature region with decreasing $T$ may be understood by considering the possible effect of a small amount of magnetic moments in the $CoO_2$ layers. The superconductivity is considered to take place in a nearly ferromagnetic phase.

Acknowledgments The authors thank Professors. Y. Ono and K. Sano for stimulating discussion. This work is supported by Grants-in-Aid for Scientific Research from the Japan Society for the Promotion of Science (JSPS) and by Grants-in-Aid on priority areas from the Ministry of Education, Culture, Sports, Science, and Technology.

Figure caption

Fig. 1    Magnetic susceptibility $\chi$ of $Na_xCoO_2 \cdot yH_2O$ measured at $H$=5G is shown against $T$. Solid and open circles correspond to the data taken under the conditions of zero-field cooling and field cooling, respectively.

Fig. 2    $^{59}$Co nuclear spin relaxation rate divided by $T$, $(1/T_1T)$ is plotted against $\ln T$ for $Na_xCoO_2 \cdot yH_2O$. Inset shows typical data of $\{m(\infty)-m(t)\}/m(\infty)$ and the solid and broken lines show the relaxation curves of the components with the shorter and longer relaxation times, $T_{1short}$ and $T_{1long}$, respectively.

Fig. 3    Magnetic susceptibility $\chi$ of $Na_xCoO_2 \cdot yH_2O$ measured with $H$= 1T is plotted against $T$. Solid and open circles are the data taken under the conditions of zero-field cooling and field cooling, respectively. The $T$-dependence of $\chi$ of $Na_{0.75}CoO_2$ is also shown by open squares. Broken and chain lines show data obtained by subtracting the Curie-Weiss components from the observed data of $Na_xCoO_2 \cdot yH_2O$ and $Na_{0.75}CoO_2$, respectively.

Fig. 4    Logarithm of the nuclear spin relaxation time $T_1$ of $^{59}$Co nuclei is plotted against $1/T$ for $Na_xCoO_2 \cdot yH_2O$. The solid line shows the relation $1/T_1T$=constant.

Fig. 5    $^{59}$Co nuclear spin relaxation rates divided by $T$, $(1/T_1T)$ of $Na_{0.75}CoO_2$ and $Na_xCoO_2 \cdot yH_2O$ are plotted against $T$ with open squares and dots, respectively.

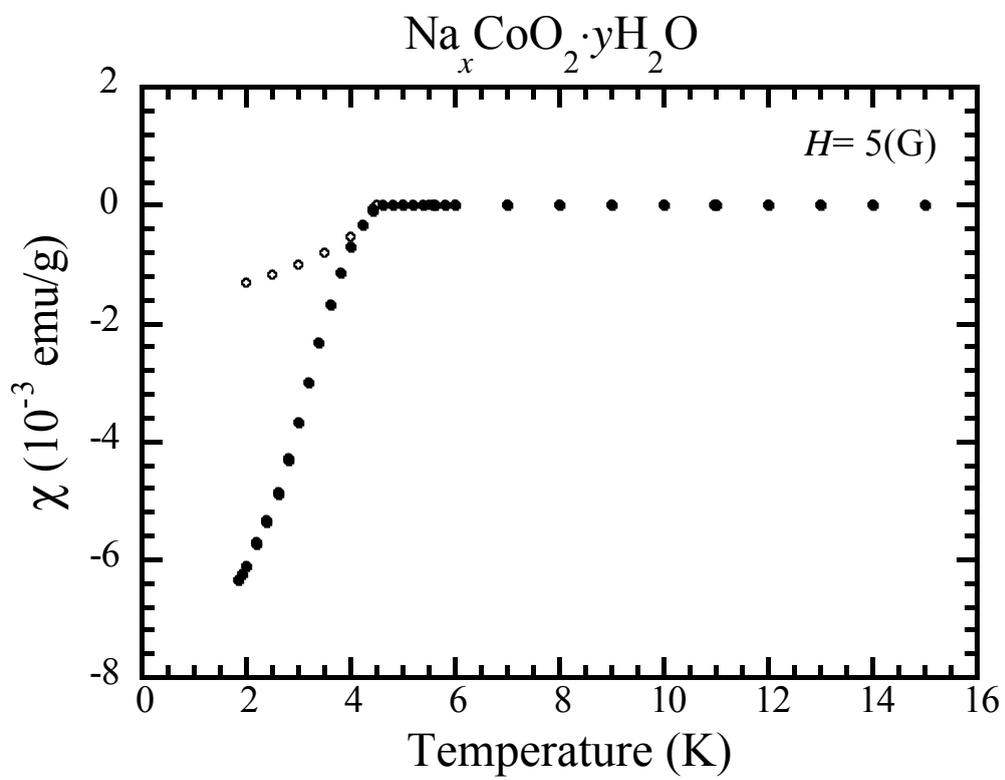

Fig. 1

Y. Kobayashi *et al.*

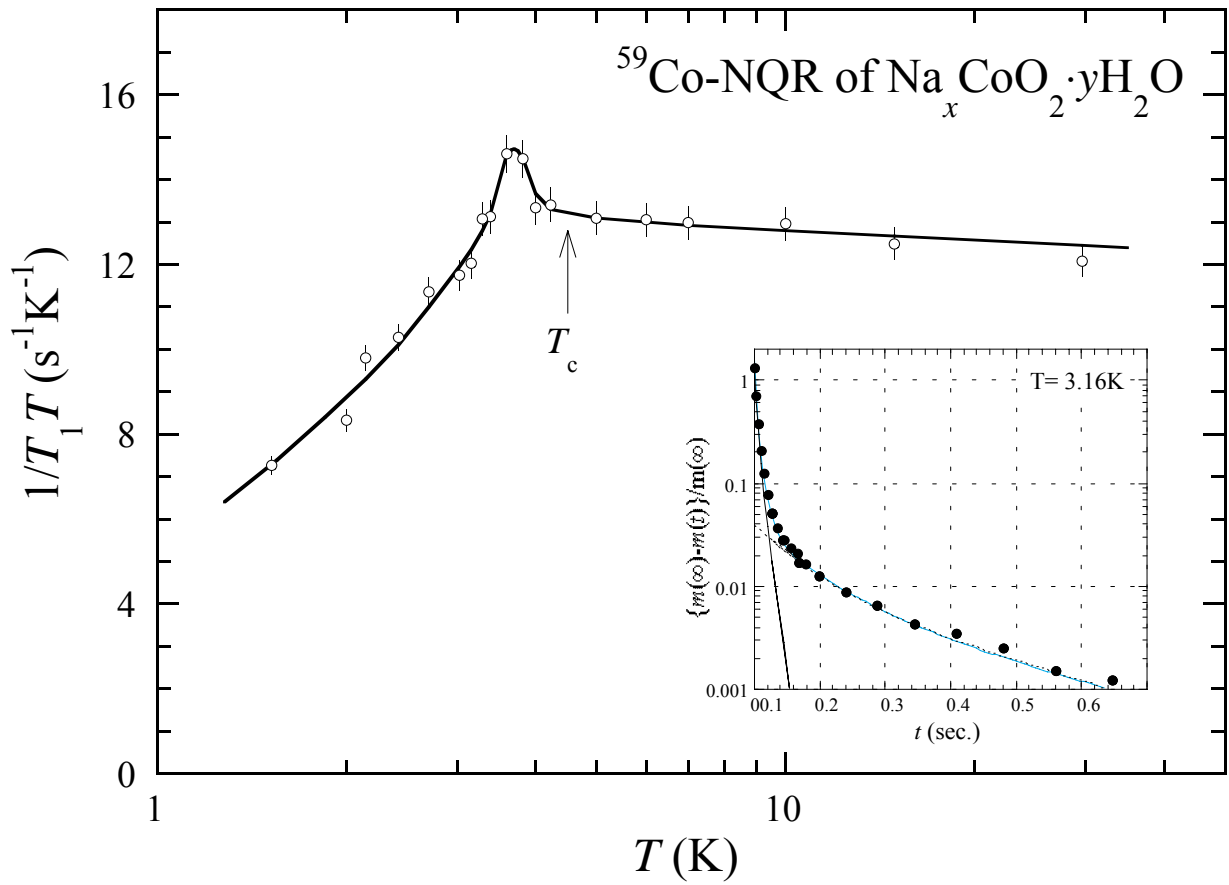

Fig. 2

Y. Kobayashi *et al*.

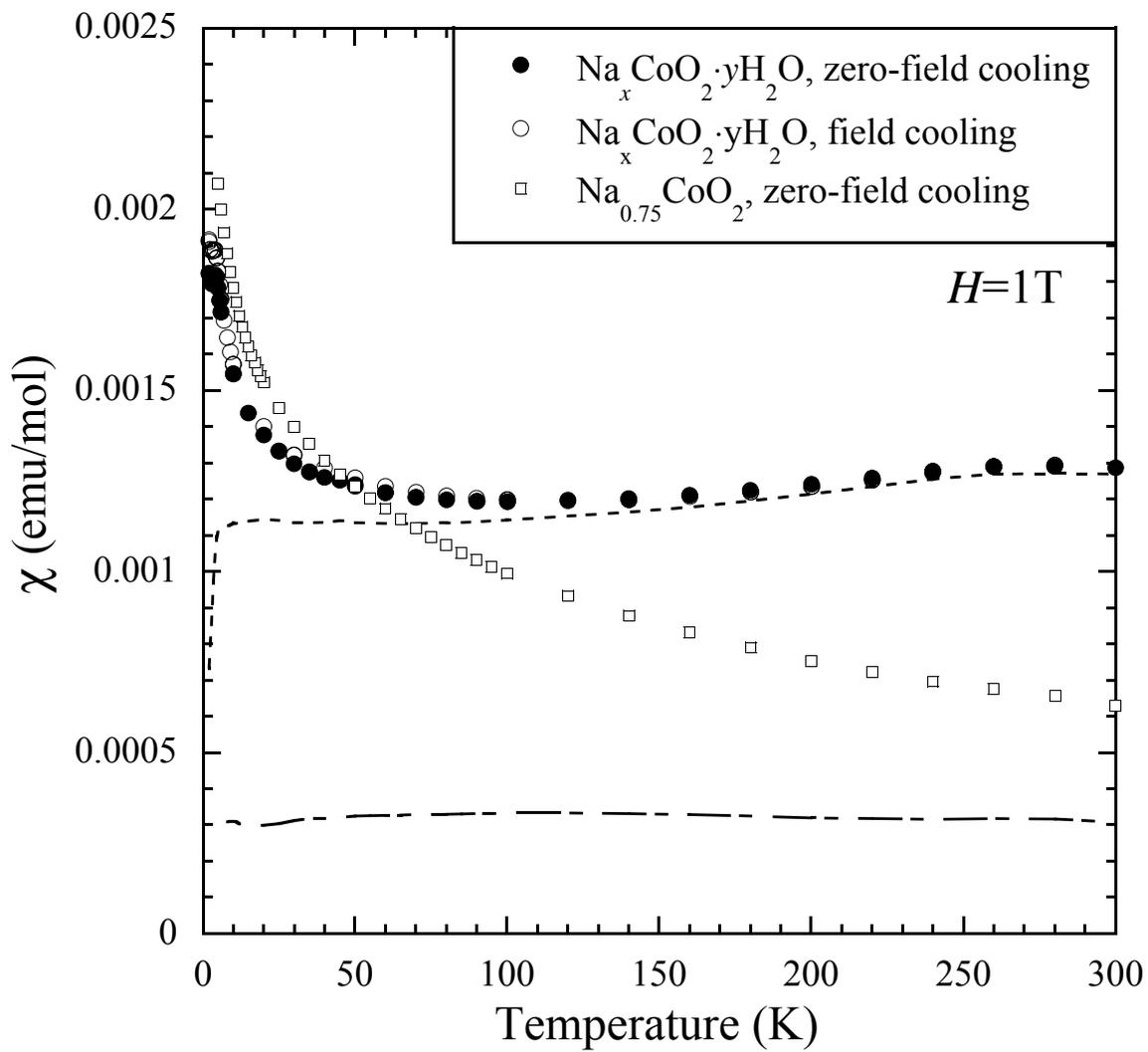

Fig. 3

Y. Kobayashi *et al*.

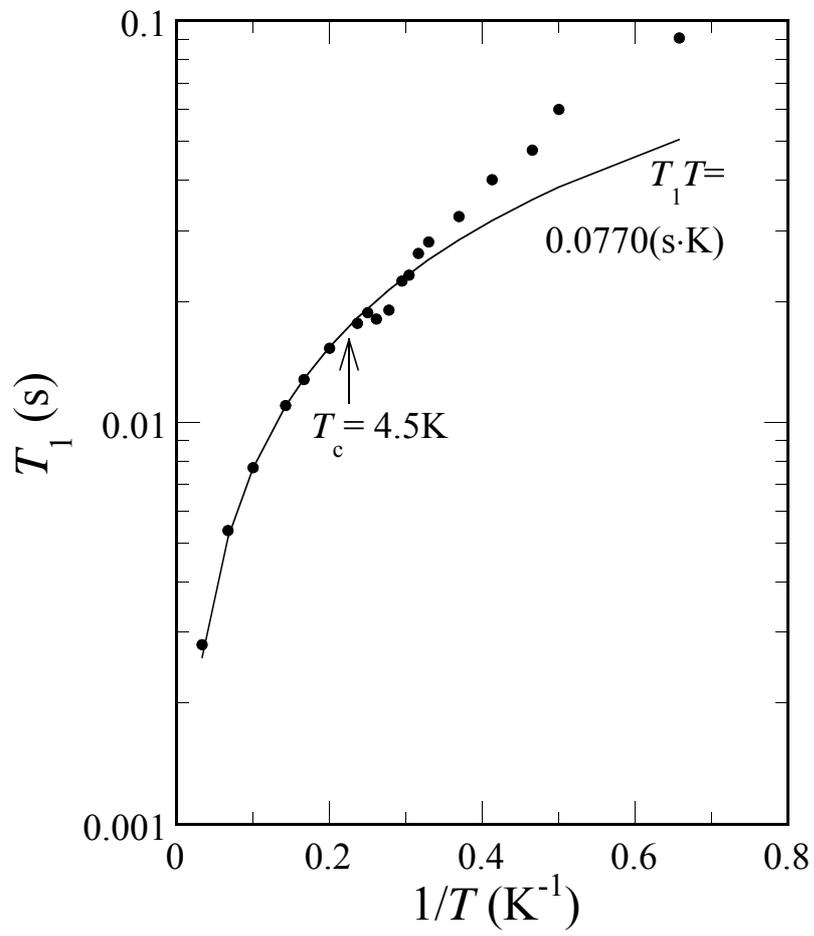

Fig. 4

Y. Kobayashi *et al*.

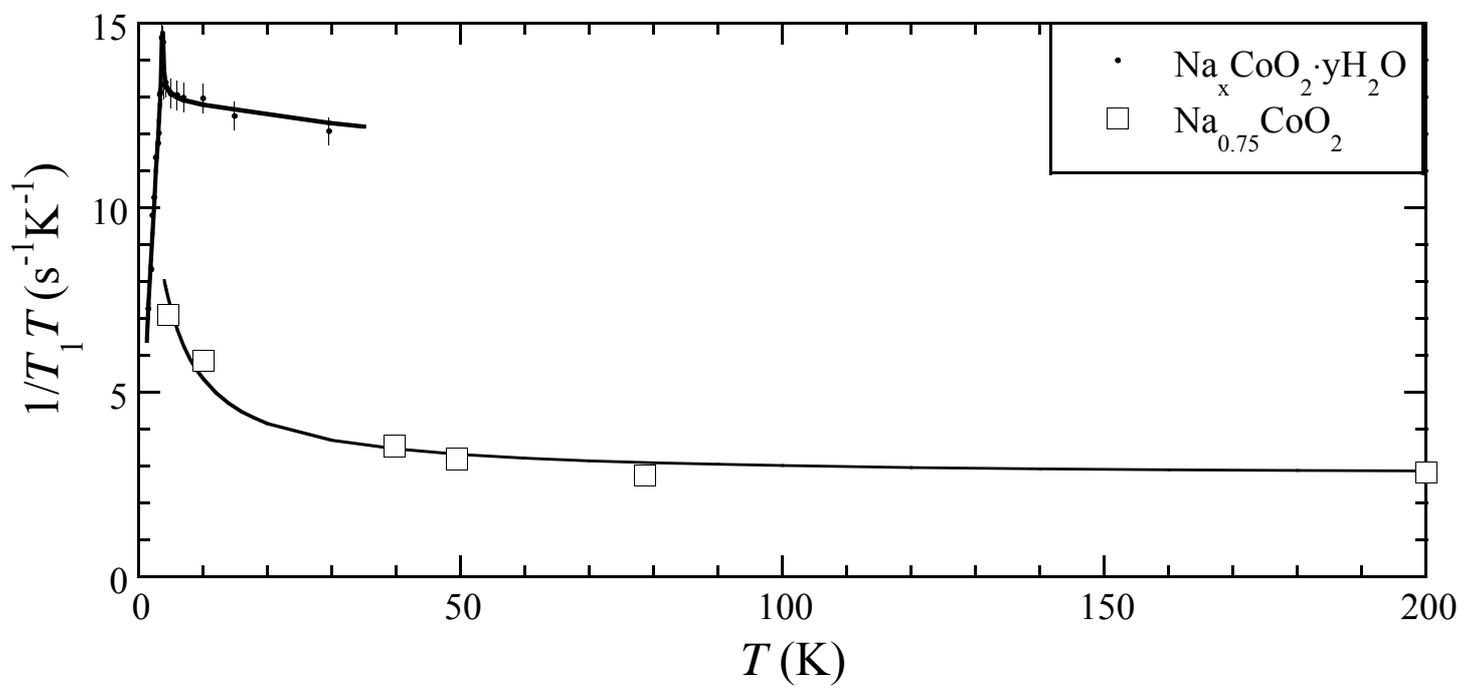

Fig. 5

Y. Kobayashi *et al.*